\documentclass[a4paper]{spie}  %>>> use for US letter paper
\usepackage{amsmath,amsfonts,amssymb}
\usepackage{graphicx}
\usepackage[colorlinks=true, allcolors=blue]{hyperref}

\title{EELT-HIRES the high-resolution spectrograph for the E-ELT}

\author[1,2]{A. Marconi}%1
\author[3]{P. Di Marcantonio}%2
\author[3]{V. D'Odorico}%3
\author[3,36]{S. Cristiani}%4
\author[4]{R. Maiolino}%5
\author[2]{E. Oliva}%6
\author[5]{L. Origlia}%7
\author[6]{M. Riva}%8
\author[7]{L. Valenziano}%9
\author[8]{F. M. Zerbi}%10
\author[9,10]{M. Abreu}%11
\author[11]{V. Adibekyan}%12
\author[12]{C. Allende Prieto}%13
\author[13]{P. J. Amado}%14
\author[14]{W. Benz}%15
\author[15]{I. Boisse}%16
\author[16]{X. Bonfils}%17
\author[17,15]{F. Bouchy}%18
\author[37]{L. Buchhave} %--- ADD SPIE mailto:buchhave@astro.ku.dk -- DONE
\author[4]{D. Buscher}%19
\author[9,10]{A. Cabral}%20
\author[18]{B. L. Canto Martins}%21
\author[19]{A. Chiavassa}%22
\author[9,10]{J. Coelho}%23
\author[21]{L. B. Christensen}%--- ADD SPIE mailto:lise@dark-cosmology.dk -- DONE
\author[11]{E. Delgado-Mena}%24
\author[18]{J. R. De Medeiros}%25
\author[20]{I. Di Varano}%26
\author[11]{P. Figueira}%27
\author[4]{M. Fisher}%28
\author[21]{J. P. U. Fynbo}%29
\author[22]{A.C.H. Glasse}%30
\author[23]{M. Haehnelt}%31
\author[4]{C. Haniff}%32
\author[21]{C. J. Hansen}%--- ADD SPIE mailto:cjhansen@dark-cosmology.dk -- DONE
\author[24]{A. Hatzes}%33
\author[25]{P. Huke}%34
\author[26]{A. J. Korn}%35
\author[18]{I. C. Le\~ao}%36
\author[27]{J. Liske}%37
\author[17]{C. Lovis}%38
\author[32]{P. Mas\l owski} % --- ADD SPIE - DONE
\author[28]{I. Matute}%39
\author[29]{R. A. McCracken}%40
\author[11]{C. J. A. P. Martins}%41
\author[11,30]{M. J. P. F. G. Monteiro}%42
\author[31]{S. Morris}%43
\author[31]{T. Morris}%44
\author[25]{H. Nicklas}%45
\author[32]{A. Niedzielski}%46
\author[10]{N. J. Nunes}%47 
\author[12]{E. Palle}%48
\author[22]{P. Parr-Burman}%49
\author[33]{V. Parro}%50
\author[23]{I. Parry}%51
\author[17]{F. Pepe}%52
\author[26]{N. Piskunov}%53
\author[4]{D. Queloz}%54
\author[34]{A. Quirrenbach}%55
\author[12]{R. Rebolo Lopez}%56
\author[25]{A. Reiners}%57
\author[29]{D. T. Reid}%58
\author[11,30]{N. Santos}%59
\author[34]{W. Seifert}%60
\author[11]{S. Sousa}%61
\author[26]{H. C. Stempels}%62
\author[20]{K. Strassmeier}%63
\author[4]{X. Sun}%64
\author[17]{S. Udry}%65
\author[35]{L. Vanzi}%66
\author[21]{M. Vestergaard}%--- ADD SPIE mailto:vester@dark-cosmology.dk - DONE
\author[20]{M. Weber}%67
\author[26]{E. Zackrisson}%68
\affil[1]{Dipartimento di Fisica e Astronomia, Universit\`a di Firenze, via G. Sansone 1, I-50019, Sesto Fiorentino (Firenze), Italy}
\affil[2]{INAF-Osservatorio Astrofisico di Arcetri, Largo E. Fermi 2, I-50125, Firenze, Italy}
\affil[3]{INAF Osservatorio Astronomico di Trieste, Via Giambattista Tiepolo 11, 34131 - Trieste Italy}
\affil[4]{Cavendish Laboratory, University of Cambridge, JJ Thomson Avenue, Cambridge CB3 0HE, UK}
\affil[5]{INAF-Osservatorio Astronomico di Bologna, Via Ranzani, 1, 40127, Bologna, Italy}
\affil[6]{INAF- Osservatorio Astronomico di Brera, Via Bianchi 46, I-23807 Merate, Italy}
\affil[7]{INAF-IASF Bologna, V.Piero Gobetti, 101 - 40129 , Bologna, Italy}
\affil[8]{Scientific Directorate of INAF, Viale del Parco Mellini 84, I-00036 Roma (Italy)}
\affil[9]{Instituto de Astrof\'isica e Ci\^{e}ncias do Espa\c{c}o, Universidade de Lisboa, Campus do Lumiar, Estrada do Pa\c{c}o do Lumiar 22, Edif. D, PT1649-038 Lisboa, Portugal}
\affil[10]{Instituto de Astrof\'isica e Ci\^{e}ncias do Espa\c{c}o, Universidade de Lisboa, Faculdade de Ci\^{e}ncias, Campo Grande, PT1749-016 Lisboa, Portugal}
\affil[11]{Instituto de Astrof\'isica e Ci\^{e}ncias do Espa\c{c}o, Universidade do Porto, CAUP, Rua das Estrelas, PT4150-762 Porto, Portugal}
\affil[12]{Instituto de Astrofisica de Canarias (IAC), C/ Via L\'actea, S/N E-38205, La Laguna (Tenerife), Espa\~na}
\affil[13]{Instituto de Astrofisica de Andalucia-CSIC Glorieta de la Astronomia s/n, 18008, Granada, Spain}
\affil[14]{Physikalisches Institut, Center for Space and Habitability, University of Berne, CH-3012 Bern, Sidlerstrasse 5, CH}
\affil[15]{Laboratoire d'Astrophysique de Marseille, CNRS, Rue Fr\'ed\'eric Joliot Curie, 13013 Marseille, France}
\affil[16]{Observatoire de Science de l'Univers de Grenoble - France}
\affil[17]{D\'epartement d'Astronomie, Universit\'e de Geneve, Chemin des Maillettes 51, Sauverny, CH-1290 Versoix, Switzerland}
\affil[18]{Board of Observational Astronomy, Federal University of Rio Grande do Norte, Campus Universit\'ario 59078-970, Natal RN, Brasil}
\affil[19]{Laboratoire Lagrange, Universit\'e C\^ote d'Azur, Observatoire de la C\^ote d'Azur, CNRS, Blvd de L'Observatoire CS34229,06004 Nice Cedex 4, France}
\affil[20]{Leibniz Institute for Astrophysics Potsdam (AIP), An der Sternwarte 16, D-14482 Potsdam, Germany}
\affil[21]{Dark Cosmology Center, Juliane Maries Vej 30, 2100 Copenhagen, Denmark}
\affil[22]{UK Astronomy Technology Centre (part of the Science and Technology Facilities Council), Blackford Hill, Edinburgh, EH9 3HJ, UK}
\affil[23]{Institute of Astronomy, Cambridge University, Madingley Rd, Cambridge, CB3 0JA, UK}
\affil[24]{Th\"uringer Landessternwarte, Sternwarte 5, 07778 Tautenburg - Germany}
\affil[25]{Institute for Astrophysics University of G\"ottingen, Friedrich-Hund-Platz 1 37077 G\"ottingen, Germany}
\affil[26]{Division of Astronomy and Space Physics, Department of Physics and Astronomy, Uppsala University, Uppsala Sweden}
\affil[27]{Hamburger Sternwarte, Universit\"at Hamburg, Gojenbergsweg 112, D-21029 Hamburg}
\affil[28]{Instituto de Astrof\'isica e Ci\^encias do Espa\c{c}o, Universidade de Lisboa, OAL, Tapada da Ajuda, PT1349-018 Lisboa, Portugal}
\affil[29]{Scottish Universities Physics Alliance (SUPA), Institute of Photonics and Quantum Sciences, School of Engineering and Physical Sciences, Heriot Watt University, Edinburgh EH14 4AS, UK}
\affil[30]{Departamento de F\'isica e Astronomia, Faculdade de Ci\^encias, Universidade do Porto, Rua do Campo Alegre, 4169-007 Porto, Portugal}
\affil[31]{Centre for Advanced Instrumentation, Department of Physics, Durham University, South Road, Durham, DH1 3LE, UK}
\affil[32]{Faculty of Physics, Astronomy and Applied Informatics, Nicolaus Copernicus University in Torun, Gagarina 11, 87-100 Torun, Poland}
\affil[33]{Instituto Mau\'a de Tecnologia, Pra\c{c}a Mau\'a, 1 - Mau\'a, S\~ao Caetano do Sul - SP, 09580-900, Brasil}
\affil[34]{Landessternwarte, Zentrum f\"ur Astronomie der Universit\"at Heidelberg, K\"onigstuhl 12, 69117 Heidelberg, Germany}
\affil[35]{Centro de Astro Ingenieria, Pontificia Universidad Catolica de Chile, Avda. Libertador Bernardo O'Higgins 340 - Santiago de Chile}
\affil[36]{INFN-National Institute for Nuclear Physics, via Valerio 2, I-34127 Trieste, Italy}
\affil[37]{Centre for Star and Planet Formation, University of Copenhagen, {\O}ster Voldgade 5-7, DK-1350 Copenhagen, Denmark}
\authorinfo{Send correspondence to A. Marconi (alessandro.marconi@unifi.it)}

\begin{document} 
\maketitle

\begin{abstract}
The first generation of E-ELT instruments will include an optical--infrared High Resolution Spectrograph, conventionally indicated as EELT-HIRES, which will be capable of providing unique breakthroughs in the fields of exoplanets, star and planet formation, physics and evolution of stars and galaxies, cosmology and fundamental physics. 
A 2-year long phase A study for EELT-HIRES has just started and will be performed by a consortium composed of institutes and organisations from Brazil, Chile, Denmark, France, Germany, Italy, Poland, Portugal, Spain, Sweden, Switzerland and United Kingdom.
In this paper we describe the science goals and the preliminary technical concept for EELT-HIRES which will be developed during the phase A, as well as its planned development and  consortium organisation during the study. 
\end{abstract}


\section{INTRODUCTION}
The exploitation of the large collecting area of the E-ELT to explore the High Spectral Resolution domain has been considered and studied since the early phases of the E-ELT program. 
Out of the Phase-A Instrument studies \cite{ramsay:2010} two have been dedicated to the spectral high resolution domain: CODEX \cite{pasquini:2010}, an ultra-stable optical (0.37-0.710 $\mu$m) high resolution ($R=120,000$) spectrograph and SIMPLE \cite{origlia:2010}, 
a Near-IR (0.84-2.5 $\mu$m) High Resolution ($R=130,000$) AO-assisted spectrograph.
When considered together, these studies demonstrated the importance of high resolution spectroscopy for the E-ELT with a \textit{simultaneous} coverage of optical and near-IR. Then, following the recommendations of the E-ELT Science Working Group and of the Scientific Technical Committee, ESO decided to include a High Resolution Spectrograph in the E-ELT instrumentation roadmap (EELT-HIRES, HIRES hereafter).
Soon after conclusion of the respective phase A studies, the CODEX and SIMPLE consortia gave birth to the HIRES initiative (http://www.hires-eelt.org) which started developing the concept of an X-Shooter-like spectrograph, but with high resolution capable of providing $R\sim 100.000$ in the 0.37-2.5 $\mu$m wavelength range. 
Following an open workshop in September 2012 the HIRES Initiative  prepared a White Paper summarizing a wide range of science cases proposed by the community \cite{maiolino:2013} and also assembled a Blue Book with a preliminary technical instrument concept. 
With the start of construction for the E-ELT, the Initiative has decided to organise itself as the HIRES Consortium and recruited additional institutes from Brazil, Denmark and Poland, which have expressed their interest in HIRES. 
The consortium answered  the Request for Information on HIRES issued by ESO in 2015, and subsequently, to the call for Phase A studies. ESO then awarded the Phase A study to the HIRES consortium and the study kick-off meeting was held at the Arcetri Observatory in Florence on March 22nd, 2016. The end of the Phase A study, including  final acceptance, is foreseen to happen by March 22nd, 2018.

\section{KEY SCIENCE CASES}

The instrument hereby proposed is conceived to be very versatile, capable of pursuing a multitude of science cases (see HIRES White Paper, Maiolino et al.~2013), but with the focus on outstanding science cases which can only be achieved through high spectral resolution with the photon collecting area provided by the E-ELT. In the following we give a summary of the main science areas:
\begin{itemize}
\item \textbf{Exoplanets} (characterization of planetary atmospheres and the detection of life signatures, exoplanets debris) 
\item \textbf{Star and planet formation} (Protoplanetary disks)
\item \textbf{Stellar physics, chemistry and astro-archaeology} (3D structure of stellar atmospheres, solar twins, stellar magnetic fields, isotope ratios and nucleosynthesis for the earliest and the latest stages of stellar evolution, chemical enrichment in the local group, extremely low metallicity stars, resolved stellar populations in extragalactic star clusters)
\item \textbf{Galaxy formation} (Population III stars, reionization, intergalactic medium, massive galaxies evolution, supermassive black holes)
\item \textbf{Fundamental Physics and Cosmology} (variation of fundamental constants, constraints on Dark Matter and Dark Energy, 
constraints on non-standard physics, Sandage Test)
\end{itemize}

\textbf{Exoplanets.} The characterization of exoplanets is one of the outstanding key science cases for HIRES. The focus will be on detecting and quantifying exoplanet atmospheres. More specifically, the unprecedented capabilities of HIRES will enable astronomers to derive the chemical composition, stratification and weather in the atmospheres of exoplanets over a wide range of planets, from Neptune-like down to Earth-like including those in the habitable zones. The ultimate goal is the detection of signatures of life. The extremely high signal-to-noise ratio required to detect the atmospheric signatures has paradoxically pushed this area into the ``photon-starved'' regime with current facilities, making the collecting area of the E-ELT essential for achieving such an ambitious goal. The key requirements for this science case are a spectral resolution $R\sim 100,000$ (primarily required to disentangle the exoplanet atmospheric features from the telluric absorption lines of our atmosphere and to increase the sensitivity by allowing the detection of narrow lines, but also to trace different layers of the atmosphere and exoplanet weather), a wide wavelength range (0.37-2.5 $\mu$m), high stability of the PSF on the detector during planetary transits and high flat-fielding accuracy. 
Within this context it is important to emphasize that HIRES will be the key instrument to follow-up the PLATO space mission, recently approved by ESA for launch in 2024 (\cite{rauer:2014}). Indeed, PLATO will provide thousands of transiting exoplanets, down to Earth-like objects, many of which will urge atmospheric characterization, but which will be feasible only through high resolution spectroscopy at extremely large telescopes \cite{udry:2014}.
A polarimetric mode would further enhance the exoplanet diagnostic capabilities of HIRES, especially for the detection of chirality as a bio-signature.
Similar capabilities (especially in the blue part of the spectrum) are required for enabling the detection of planetary debris on the surface of white dwarfs, which is an alternative and exciting technique to trace the composition of exoplanets. 
In the ELT era, radial velocity studies will not only focus on the detection of exo-Earths, but also on the measurement of the weak and rare time-limited Rossiter-McLaughlin effect to determine the spin-orbit obliquity of terrestrial planets. The latter will require a stability of the spectrometer of 10 cm s$^{-1}$.

\textbf{Star and planet formation.} HIRES will also provide the capability of revealing the dynamics, chemistry, and physical conditions of the inner-most regions of stellar accretion disks as well as protoplanetary disks of young stellar objects. Therefore, it provides unprecedented constraints on the physics of star formation, jet launching mechanisms and planet formation in general. These observations will be highly synergic with those at other wavelengths from facilities like ALMA. To achieve these goals in the near-IR, the instrument's high spectral resolution ($R\sim 100,000$) should be accompanied with spatially resolved information (possibly with an Integral Field Unit - IFU - mode) at the diffraction limit of the E-ELT. This science case would also benefit from a polarimetric mode, which would provide information on the magnetic field in the funnelling region of the accretion disk.

\textbf{Stellar physics, chemistry and astro-archaeology.} Stars are far from being understood. In particular, dynamical phases of stellar evolution are not well explored. Although new simulations of stellar convection, differential rotation, stellar magnetic activity, mass loss, and interior mixing, as well as star formation and the interaction of stars with protoplanetary disks and planets are being developed, these efforts are to a considerable degree dependent on high simplifications and unknown initial conditions. The only way to remedy this situation is to guide theoretical developments with better observations. Among the most demanding observable is the magnetic field of a star and its role for stellar evolution and cosmic-ray shielding for its planets. Was the civilization on our planet only possible because the Sun has a magnetic field of just the right strength? High-resolution spectral and polarimetric observations of solar twins at the age of the Sun and beyond will clearly contribute to the answer. The detection of biomarkers is an already ongoing race. There is also the interesting new possibility of detecting the presence of terrestrial planets from stellar spectroscopy alone, based on the unusual volatile-to-refractory ratio of the Sun.
As for galactic archaeology, HIRES will deliver for the very first time the high resolution and high quality ($S/N>100$) required to trace in detail the chemical enrichment pattern of solar-type and cooler dwarf stars out to distances of several kpc, therefore sampling most of the Galactic disk and bulge. HIRES will also enable astronomers to chemically characterize sub-giants and red giants in the outer Galactic halo and in neighbouring dwarf galaxies. With a spectral resolution of $R\sim 100,000$ and a broad spectral coverage (0.37-2.5 $\mu$m), the detailed chemical mapping of numerous elements and isotopes will reveal the origin and the formation history of ancient stars. This will be crucial for the extremely low metallicity stars, whose photospheres may trace the chemical abundances resulting from the enrichment of the first population of stars (Pop-III). 

\textbf{Galaxy formation.} In the context of galaxy formation and cosmology, one of the most exciting prospects for HIRES is the detection of elements synthesized by the first stars in the Universe. The direct detection of Pop-III stars (through the associated UV nebular features, e.g., HeII emission), in the early Universe, is probably out of reach also for JWST, and subject to interpretation ambiguities. HIRES will probably be the first facility that could unambiguously detect the fingerprint of Pop-III stars by measuring the chemical enrichment typical of this population in the Inter-Galactic (IGM) and Inter-Stellar Medium (ISM) in the foreground of Quasars, GRBs and Super-Luminous Supernovae at high redshift, probing in this way the epoch of reionization. These observations will reveal the nature and physical properties of the first stars that populated the Universe, indeed different potential Pop-III progenitors are expected to result in specific chemical enrichment patterns.
The high spectral resolution of HIRES will also allow astronomers to trace in detail the history of the reionization process of the Universe and the subsequent thermal history of the IGM. This will in turn provide crucial information on the classes of sources responsible for the reionization of the Universe (in particular young stars, Pop-III or AGNs). To reach these exciting science goals with HIRES requires a spectral resolution $R>50,000$ and a spectral coverage extending from about 4000\,\AA\ to 2.5 $\mu$m.
If enabled with some multiplexing capability (5-10 objects), HIRES will also be able to obtain a three-dimensional map of the cosmic web of the IGM at high redshift, by probing absorption systems towards multiple lines of sight on scales of a few arcminutes and by targeting star forming galaxies. This technique has already been successfully tested with existing telescopes. The much higher sensitivity of HIRES will enable astronomers to obtain a far more complete tomographic map of the IGM by sampling a much higher density of lines of sight (by accessing fainter galaxies) and down to much lower intervening absorption columns. Most importantly, if the simultaneous wavelength coverage extends from 4000\AA\ to 2.5 $\mu$m, HIRES will have the exceptional capability of obtaining a three-dimensional map of the distribution of metals in the IGM, which would be a unique probe of the enrichment process of the Universe. 
When equipped with an IFU sampling the ELT diffraction limit, HIRES, with its high spectral resolution ($R\sim 100,000$) will be the only tool to measure the low mass end of supermassive black hole in galactic nuclei, down to $\sim 10^{4}-10^{5} M_{\odot}$. By investigating whether the galaxy-black hole scaling relations hold down to very low masses, or there are significant deviations, it will be possible to test and discriminate between different galaxy-black hole co-evolutionary theories. Most importantly, low mass black holes bear the signature of primordial black hole seeds, hence by probing their mass distribution and relation with their host galaxies, it will be possible to test different scenarios that have been proposed for the primordial formation of black holes.

\textbf{Fundamental Physics and Cosmology.} Perhaps most exciting, HIRES will be an instrument capable of addressing issues that go beyond the limited field of Astronomy, breaking into the domain of ``fundamental physics''. In particular, HIRES will provide the most accurate tests of the stability of the fundamental constants of nature such the fine structure constant $\alpha$ and competitive constraints on dynamical dark energy and Weak Equivalence Principle violations. Similarly, competitive tests can be made for the proton-to-electron mass ratio $\mu$. These measurements require that HIRES has a spectral resolution $R\sim 100,000$ and high efficiency in the blue part of the spectrum.
HIRES will also deliver the most accurate measurements of the CMB temperature at high redshift (using CO measurements, which are S/N limited) and of the deuterium abundance, both of which will provide stringent constraints on models of non-standard physics.
The exquisite S/N and high spectral resolution delivered by HIRES will also enable a statistically robust study of the profile of the Ly$\alpha$ forest, which among other things will provide extremely tight constraints on the nature of Dark Matter, and in particular any putative contribution by warm dark matter (e.g. gravitinos).
Many of the spectra gathered for the above purposes will also contribute to the first measurement of the redshift drift-rate $dz/dt$ deep in the matter dominated era. This is a a unique, direct, non-geometric and completely model-independent measurement of the Universe's expansion history (the ÒSandage testÓ). This should be regarded as the beginning of a legacy experiment. In addition to high-spectral resolution ($R\sim 100,000$), this measurement requires excellent wavelength calibration with a precision of about 70 cm s$^{-1}$ (which can be achieved with laser comb technology and high fibre scrambling efficiency) and with a stability of the accuracy of the calibration, of the order of 2 cm  s$^{-1}$, over the duration of the lifetime of the instrument.

\section{SCIENCE REQUIREMENTS AND COMPARISON WITH ESO TOP LEVEL REQUIREMENTSs}

In summary, the various science cases result in the following set of requirements:
\begin{itemize}
\item A primary high-resolution observing mode with $R\sim 100,000$ and a simultaneous wavelength range $0.37-2.5$ $\mu$m (although the extension to 0.33 $\mu$m is desirable for some cases).
\item For most science cases a stability of about 10 cm s$^{-1}$ and an accuracy of the relative wavelength calibration of 1 m s$^{-1}$ are sufficient. The exoplanet radial velocity cases also require a wavelength accuracy down to 10 cm s$^{-1}$.
\item The Sandage test requires a stability as good as 2 cm s$^{-1}$ over the duration of a night and also an absolute wavelength calibration of 2 cm s$^{-1}$. These numbers are at the date considered as desirable goals more than design drivers.
\item The science cases of mapping the  large scale structure matter distribution, galaxy evolution and extragalactic star clusters would greatly benefit from having, within the same wide spectral coverage (0.37-2.5 $\mu$m), a moderate multiplexing capability (5-10 objects within a FoV of a few arcminutes) with a moderate spectral resolution mode ($R\sim 10,000-50,000$).
\item Most of the extragalactic, high-z science cases require an accurate subtraction of the sky background, to better than 1\%.
\end{itemize}

Table \ref{tab:TLR} provides a summary of the top level requirements that, according to the consortium, would enable HIRES to achieve most of the science goals outlined above and in the White Paper. These requirements are compared with the Top Level Requirements (TLR) issued by ESO in document ESO-204697, ÒTop Level Requirement for ELT-HIRESÓ.

It is worth noticing that the call for phase A studies issued by ESO requires a baseline design with a cost cap of 18 MEuros. This is well below the rough cost estimates made in the  early phases by the HIRES Initiative  for an instrument capable of fulfilling all TLRs. Therefore the selection of a subset of TLRs to drive the instrument design must be performed at the beginning of the phase A study (see below).
%\begin{table}[t]
%\includegraphics[width=0.7\linewidth]{table1.pdf}
%  \begin{minipage}[b]{0.28\textwidth}   \caption{\label{tab:TLR}Top Level Requirements prepared by the HIRES Consortium and comparison with those issued by ESO (ESO-204697).}  \end{minipage}
%\end{table}
\begin{table}[t]
   \caption{\label{tab:TLR}Top Level Requirements prepared by the HIRES Consortium and comparison with those issued by ESO (ESO-204697).} 
   \begin{center}\includegraphics[width=0.7\linewidth]{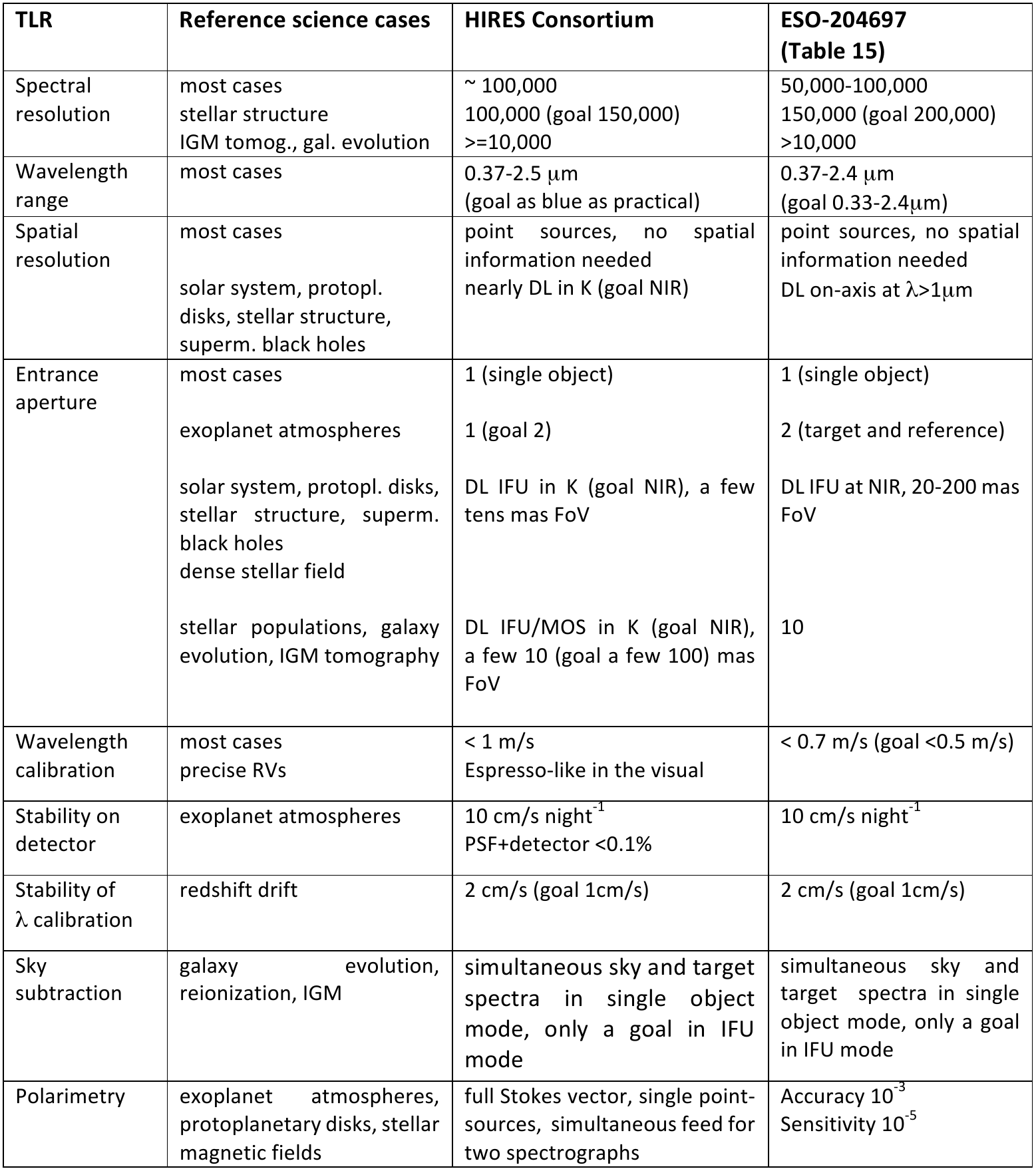}\end{center}
\end{table}

\section{POSSIBLE INSTRUMENT CONCEPT TO FULFIL THE SCIENCE REQUIREMENTS}
\label{sec:sections}
%\begin{figure} [t]
%%\begin{tabular}{c} 
%   \includegraphics[width=0.7\linewidth]{figure1.pdf} \hfill
%  \begin{minipage}[b]{0.28\textwidth}   \caption[example]{\label{fig:hires}Possible HIRES architecture with wavelength splitting in four spectrographs. Wavelength splitting is indicated only for illustrative purposes and will be the results of the trade-off analysis conducted during the Phase A study.  } \end{minipage}
%%\end{tabular}
%\end{figure} 
\begin{figure} [t]
\begin{center}
   \includegraphics[width=0.6\linewidth]{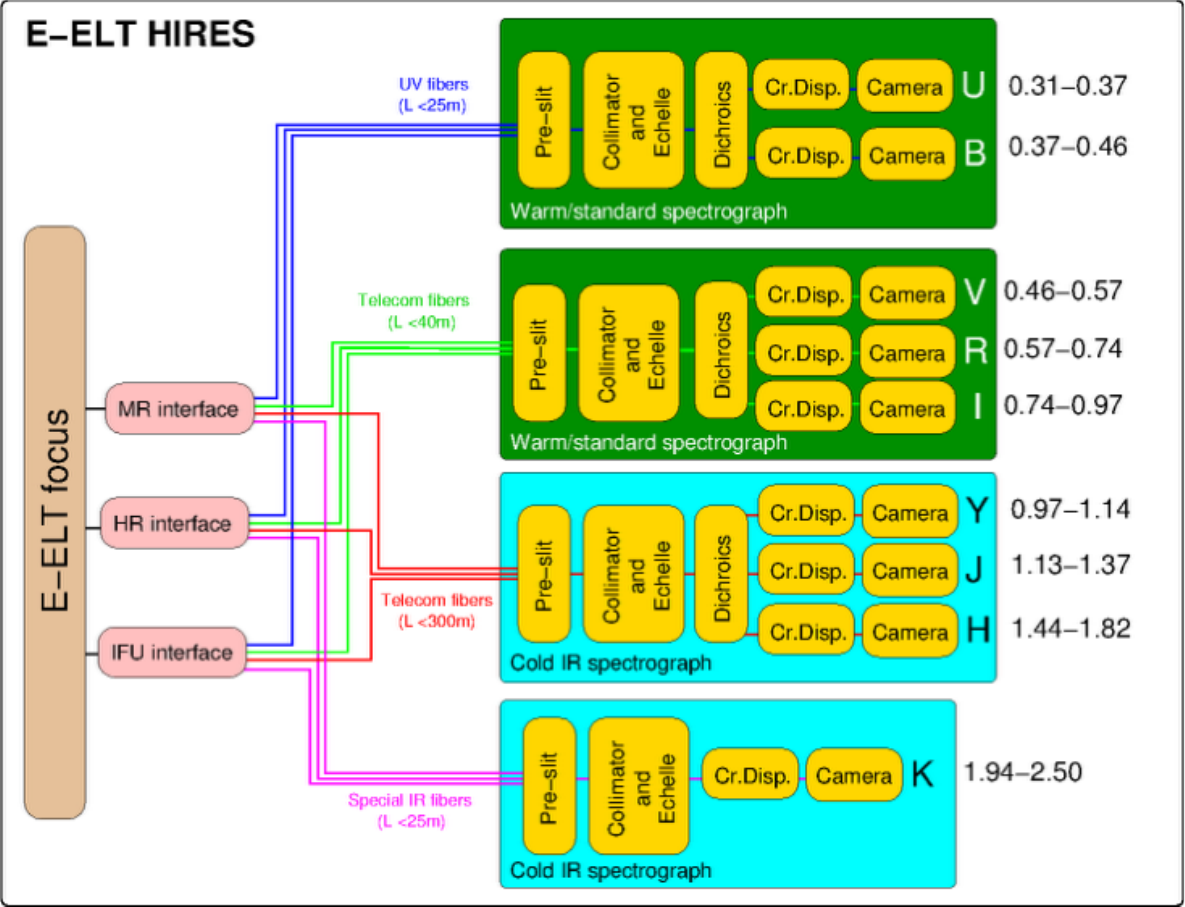} \hfill
  \caption[example]{\label{fig:hires}Possible HIRES architecture with wavelength splitting in four spectrographs. Wavelength splitting is indicated only for illustrative purposes and will be the results of the trade-off analysis conducted during the Phase A study.  } 
  \end{center}
\end{figure}

In this section, we describe a possible concept for a modular instrument in line with all requirements presented in table \ref{tab:TLR}. This concept is based on the preliminary technical study conducted by the HIRES initiative. The modular nature of this design offers the advantage of easy, non-destructive de-scoping of the instrument to match the cost-cap imposed by ESO: one or more modules could be built within the 18 MEuros cost cap with the others waiting for more money to become available.

The design of an instrument operating from the blue to the K band requires different detector technologies to be encompassed in the design not to mention the different required temperature regimes, e.g.~cryogenic environment beyond 1 micron.  A modular fibre-fed cross-dispersed echelle spectrograph is considered as a promising concept to be deeply studied.
Such concept has been investigated by the HIRES consortium and is summarised in Figure \ref{fig:hires}. It is described in detail in the Blue Book. The light from the telescope 
is split via dichroics into N wavelength channels. Each wavelength channel includes several interfaces that feed, through separate groups of fibres, a dedicated spectrograph module. Each interface and fibre bundle corresponds to an observing mode. Based on its preliminary analysis, we consider the four spectrograph modules shown in Figure \ref{fig:hires} as our starting point. 
The wavelength separation between the modules is influenced, among many other parameters, by the optical transparency of the different types of fibres available on the market. Therefore, the different modules could be positioned at different distances from the telescope focal plane depending of how much light loss one is willing to take. The flexibility of detaching the modules from the telescope, allowed by the fibres, is an important advantage of this design.
All spectrograph modules have a fixed configuration, i.e., no moving parts inside the spectrograph. They consist of a series of parallel entrance slits, each generated from a separate set of fibres that, in turn, determines the observing mode. A definition of the baseline observing modes is foreseen for this phase-A study.
An integral part of the instrument concept is its data flow and control. The preliminary data flow concept investigated by consortium follows an end-to-end scheme of operations with the final goal of providing the astronomer with high-level scientific products as complete and precise as possible in a short time after the end of an observation. This will be further investigated during the phase A study in several dedicated work packages, from data reduction and data analysis to low level instrument control.
Table \ref{tab:compliance} (top) summarises the compliances of the preliminary design with the TLRs issued by ESO and shows the existing instruments which justify the fulfilment of the TLRs. Table \ref{tab:compliance} (bottom) shows the instrument heritage at a subsystem level. Table \ref{tab:compliance} clearly indicate that there are almost no technical challenges for this design, and that this is based on existing instruments all built by consortium members.

In conclusion, considering the top level requirements set by Table \ref{tab:TLR}  and the ESO Document ESO-204697 ÒTop Level Requirement for ELT-HIRESÓ, a preliminary evaluation of a possible concept of a modular fibre-fed cross-dispersed echelle spectrograph drives to the following conclusions that will have to be validated by the Phase A study.
An Instrument able to provide high-resolution spectroscopy (100,000) over a wide wavelength range (0.37-2.5 $\mu$m) can be built with currently available technology. No obvious show-stoppers were identified but some R\&D is foreseen to maximise performance and/or reduce the cost. The project can be developed with a staged timeline for the various sub-systems, thereby adjusting to the available funds.  
Furthermore, the proposed concept will comprise the following advantages:
\begin{itemize}
\item Flexible module location, i.e., the location of the different modules may adapt to the development of the telescope.
\item Fibre-fed instrument designed for seeing-limited observations: its main performances will not depend on external AO although AO will allow additional observing modes (e.g. IFU).
\item Insensitivity of HIRES PSF to variations of pupil illumination (due, e.g., to missing segments from the primary mirror).
\item High dynamic range making it possible to reach a very high S/N in a single exposure.
\end{itemize}

\begin{table}[ht]
\caption{\label{tab:compliance} Top Level Requirements compliance and past/existing projects to justify compliance (top). Same as top table but at a subsystem level (bottom).} 
\begin{center}
\includegraphics[width=0.7\linewidth]{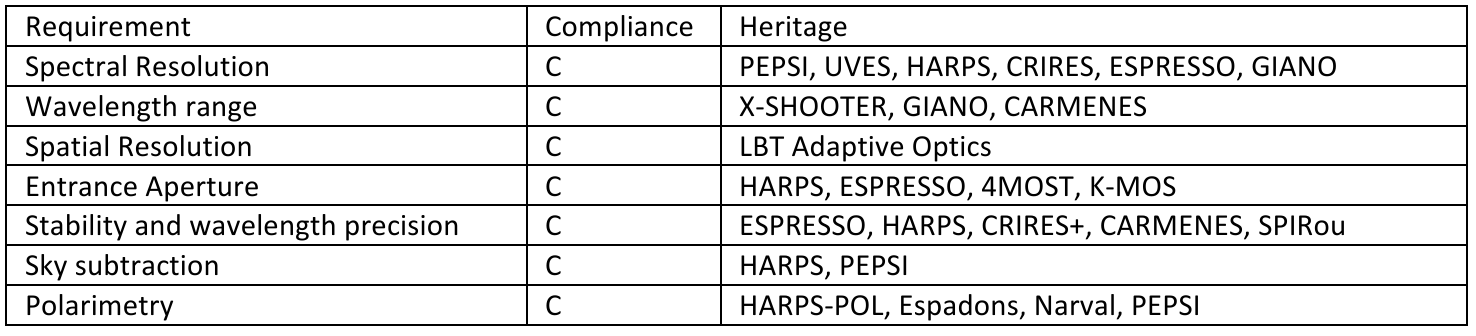}
\vskip 0.5cm
\includegraphics[width=0.45\linewidth]{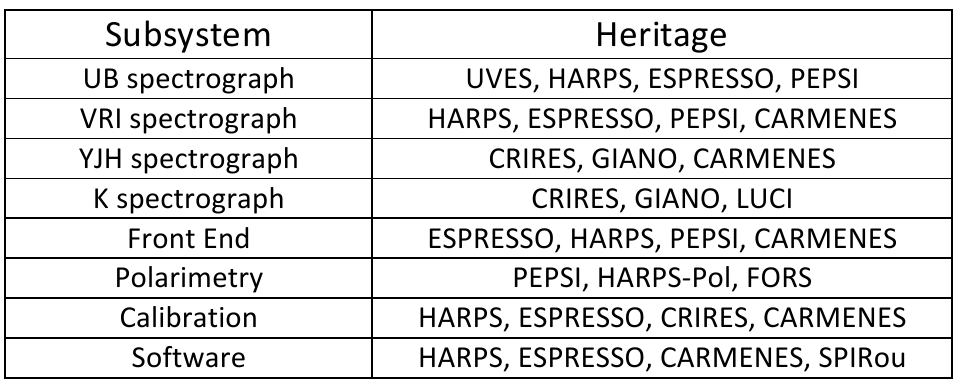}
\end{center}
\end{table}

\section{\normalsize Consortium organization}
\begin{table}[t]
\caption{\label{tab:composition} Consortium composition. The first column indicates the member countries and the corresponding co-I, member of the executive board. The second column indicates the coordinating institutes (consortium members) while the third column lists the associated members from each country.} 
\begin{center}
\begin{tabular}{|p{0.2\linewidth}|p{0.35\linewidth}|p{0.35\linewidth}|}
\hline
\textbf{Country} & \textbf{Coordinating Institution} & \textbf{Other Consortium Members} \\
\hline
\textbf{Brazil \newline (J. Renan de Medeiros)} &
\small Board of Observational Astronomy, Federal University of Rio Grande do Norte &
\small Instituto Mau\'a de Tecnologia \\
\hline
\textbf{Chile \newline(L.~Vanzi)}&
\small Pontificia Universidad Catolica de Chile, Centre of Astro Engineering &
\small Universidad de Chile, Department of Astronomy; Universidad de Concepcion, Center of Astronomical Instrumentation; Universidad de Antofagasta \\
\hline
\textbf{Denmark \newline(J.~Fynbo)}&
\small Niels Bohr Institute, University of Copenhagen & 
\small Department of Physics and Astronomy, Aarhus University\\
\hline
\textbf{France\newline(I.~Boisse)}&
\small Laboratoire d'Astrophysique de Marseille &
\small Institut de Plan\'etologie et d'Astrophysique de Grenoble, Universit\'e Grenoble Alpes; Laboratoire Lagrange, Observatoire de la C\^ote d'Azur;
Observatoire de Haute Provence, CNRS, Aix Marseille Universit\'e, Institut Pyth\'eas\\ 
\hline
\textbf{Germany \newline(K.~Strassmeier)}&
\small Leibniz-Institut f\"{u}r Astrophysik Potsdam (AIP) &
\small  Institut f\"{u}r Astrophysik, Universit\"{a}t G\"ottingen (IAG); Zentrum f\"{u}r Astronomie Heidelberg, Landessternwarte (ZAH); Th\"{u}ringer Landesternwarte Tautenburg (TLS); Hamburger Sternwarte, Universit\"{a}t Hamburg (HS)\\
\hline
\textbf{Italy\newline (A.~Marconi - PI)}&
\small Istituto Nazionale di Astrofisica (INAF) - Lead Technical Institute&\\
\hline
\textbf{Poland \newline(A.~Niedzielski)}&
\small Faculty of Physics, Astronomy and Applied Informatics, Nicolaus Copernicus University in Torun&\\
\hline
\textbf{Portugal \newline(N.~Santos)}&
\small Instituto de Astrof\'isica e Ci\^encias do Espa\c{c}o (IA) at Centro de Investiga\c{c}‹o em Astronomia/Astrof\'isica da Universidade do Porto (CAUP) &
\small  Instituto de Astrof\'isica e Ci\^{e}ncias do Espa\c{c}o (IA) at Faculdade de Ci\^{e}ncias da Universidade de Lisboa\\
\hline
\textbf{Spain \newline(R.~Rebolo)}&
\small Instituto de Astrof\'isica de Canarias & 
\small Instituto de Astrof\'isica de Andaluc\'ia-CSIC; Centro de Astrobiolog\'ia\\
\hline
\textbf{Sweden\newline (N.~Piskunov)}&
\small Dept. of Physics and Astronomy, Uppsala University &\\
\hline
\textbf{Switzerland\newline (F.~Pepe)}&
\small D\'epartement d'Astronomie, Observatoire de Sauverny, Universit\'e de Gen\`eve & 
\small Universit\"{a}t Bern, Physikalische Institut\\
\hline
\textbf{United Kingdom\newline (M.~Haehnelt)}&
\small Science and Technology Facilities Council & 
\small Cavendish Laboratory \& Institute of Astronomy, University of Cambridge; UK Astronomy Technology Centre; Centre for Advanced Instrumentation - Durham University;
Institute of Photonics and Quantum Sciences, Heriot-Watt University\\
\hline
\end{tabular}
\end{center}
\end{table}
The Consortium is composed of institutes from Brazil, Chile, Denmark, France, Germany, Italy, Poland, Portugal, Spain, Sweden, Switzerland and United Kingdom (table \ref{tab:composition}). 
For each country, one institute ( ``Coordinating Institution'') coordinates the contributions from all other institutes of that country (``Other Consortium Members'').  The Italian National Institute for Astrophysics (INAF) is the lead technical Institute and Alessandro Marconi, from the University of Florence and INAF, is the PI of the consortium. 
\begin{figure} [ht]
\begin{center}
\begin{tabular}{c} 
   \includegraphics[width=0.8\linewidth]{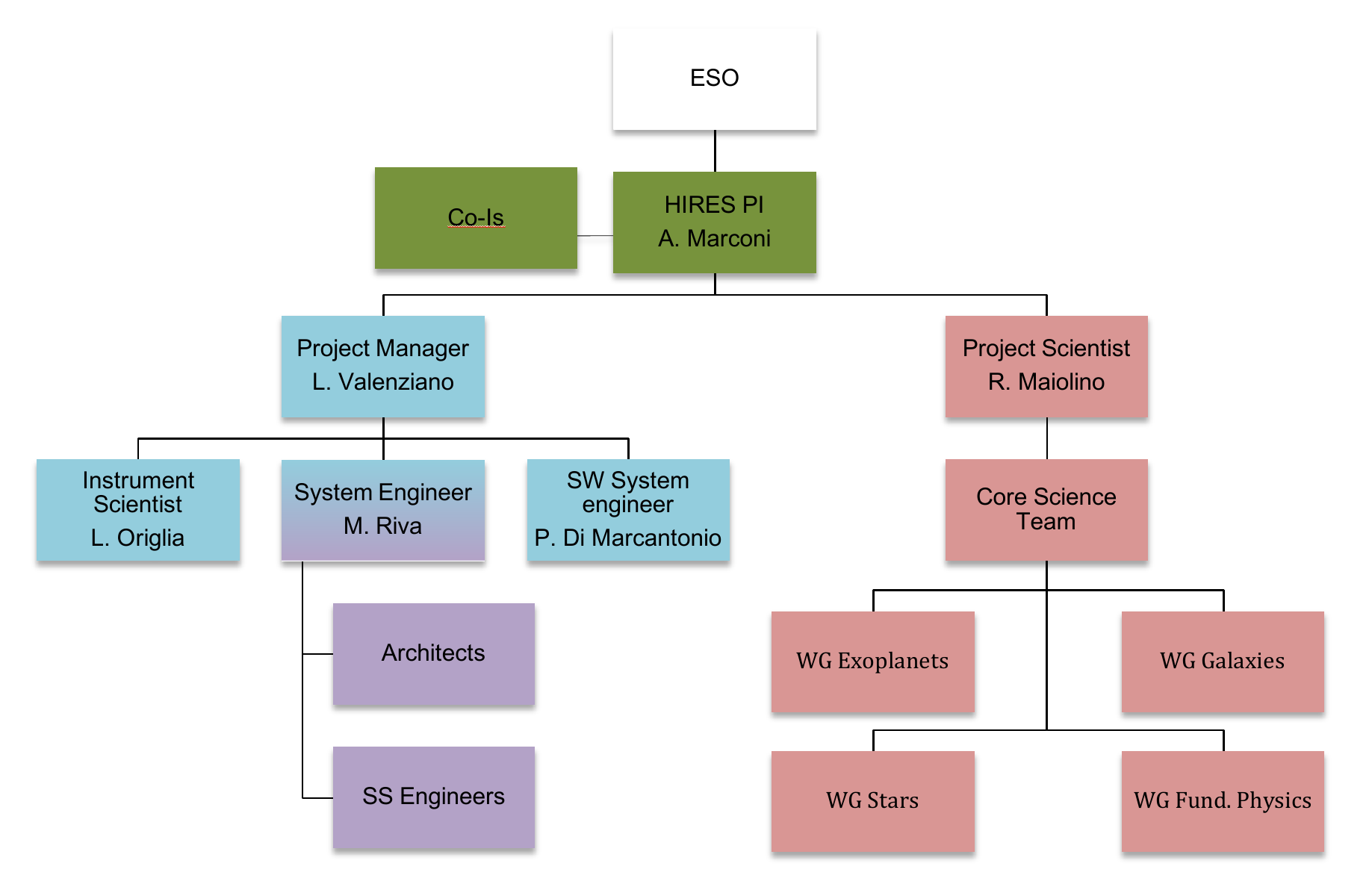}
\end{tabular}
\end{center}
\caption[example]{\label{fig:structure} Consortium structure. Different colors denote different areas of activity (red - science team, green - executive board, blu - project office, magenta -  technical team)}
\end{figure} 

The consortium structure is schematically represented in Figure \ref{fig:structure}. The HIRES project is coordinated by the Principal Investigator (PI; A. Marconi) who is also the contact person between ESO and the HIRES consortium.  The Board of co-Investigators (co-Is) is composed of one representative per country (see Table \ref{tab:composition}; all decisions concerning the overall scientific performance and exploitation of the instrument and all matters concerning the organization of the consortium are taken by the PI and the Board. During the meetings with the Board the PI is assisted by the Project Scientist (PS; R, Maiolino) and by the Project Manager (PM; L. Valenziano). 
The PS chairs the Science Advisory Team (SAT) and is responsible for finalising the top level requirements and providing the link between the science and technical team.
The SAT is composed of a science team at large and of a core science team. The science team at large is composed of 4 working groups, each with a coordinator and a deputy.
\begin{itemize}
\item Exoplanets and  Circumstellar Disks (Coordinator E. Palle, Deputy C. Lovis) 
\item Stars and Stellar Populations (C. Allende, A. Korn) 
\item Formation and Evolution of Galaxies and Intergalactic Medium (V. D'Odorico, E. Zackrisson)
\item Cosmology and Fundamental Physics (J. Liske, C. Martins)
\end{itemize}
Each working group is composed of about 15 members beyond coordinator and deputy. The PS, as chair, the coordinators and deputies of each working group constitute the core science team. The science team is responsible for all matters that concern the science cases for the instrument, for instance for the prioritisation of the key science cases that will allow to help defining the baseline design of the instrument.

The Project Office (PO) is coordinated by the PM and includes the Instrument Scientist (IS; Livia Origlia), the System Engineer (SE; M. Riva), the Software System Engineer (SSE; P. Di Marcantonio). The project office coordinates the activities of the various Work Packages (WP) which include:
Visible sub-system (i.e. the one or more visible spectrographs), Infrared sub-system (i.e. the one or more near-infrared spectrographs), Front End, Calibration, Software, Fibre Link, Polarimeter, Multifunctional sub-system (i.e. AO, IFU and MOS observing modes, etc.).

\begin{figure} [ht]
\begin{center}
\begin{tabular}{c} 
   \includegraphics[width=0.6\linewidth]{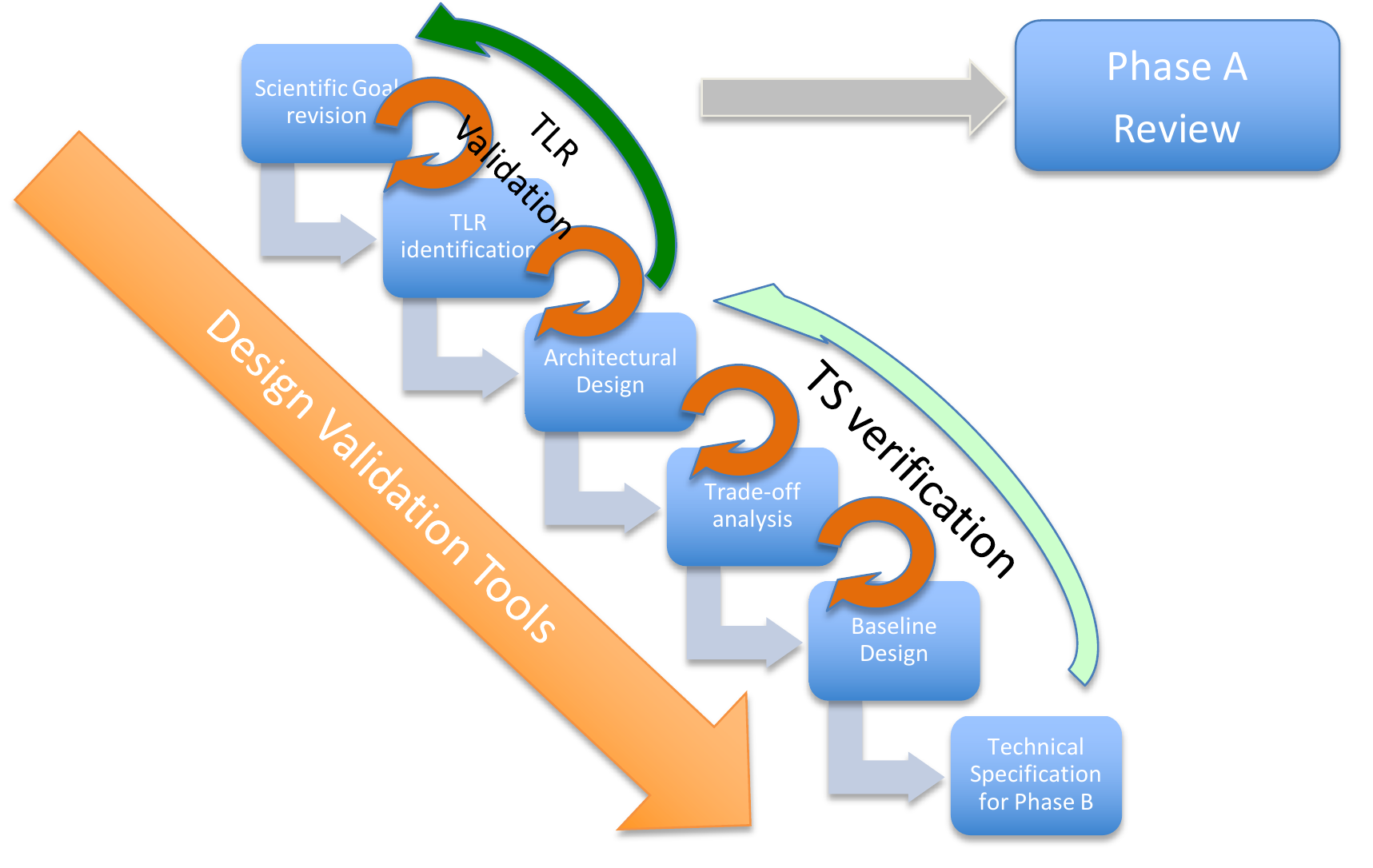}
\end{tabular}
\end{center}
\caption[example]{\label{fig:plan}Development of the phase A study.  }
\end{figure} 
\section{DEVELOPMENT OF THE PHASE A STUDY}

The phase A study will be developed according to the  steps indicated in figure \ref{fig:plan} and described below.
\begin{itemize}
\item Science Goals and Priorities. The SAT identifies the top priority science cases for the instrument, as a result of the work done in each  working group which is then merged by the core SAT. 
\item Instruments TLRs definition \& Technical specifications. The PS, with the support of the SAT, identifies the science requirements and defines the TLRs which must be a subset of those provided by ESO. The technical team, in collaboration with the PS, prepares technical specifications based on the TLRs.
\item Architectural design \& Trade-off analysis. The technical team prepares a straw man design. The technical team identifies the different possible configurations that match the TLRs. The technical team also identifies possible extensions to each configuration which can help a better matching of the TLR (i.e. additional parts likely requiring extra funding with respect to the cost cap).
\item Baseline design. The technical team further iterates and extracts a first baseline design that also matches the cost cap by ESO and provides a first evaluation of technical feasibility and risks. 
\item Technical Specification verification. The Technical team, in collaboration with the PS, verifies the match between the Technical Specification and the baseline instrument design. 
\item TLR validation. The PS informs the SAT which evaluates the impact of the proposed baseline instrument design on the top science cases. The SAT identifies pros and cons for the top science cases with a given configuration (e.g. dropping a given module or including another one). The SAT also studies the feasibility of the science cases with lower priority.
\item The PS revises the TLRs providing input to the technical team for a revision of the baseline design. A new baseline design is produced and the process is iterated to a satisfactory convergence.
\end{itemize}

\section{GENERAL REMARKS}
The preliminary design based on four modular fibre-fed cross-dispersed echelle spectrographs promises to match all TLRs issued by ESO but its cost estimate is likely well above
the envisioned cost-cap of 18 MEuros. One possibility would be to find a much cheaper design per module spectrograph but, considering that each of the four modules corresponds to a VLT instrument like ESPRESSO this possibility is highly unlikely. Therefore, the only option is a de-scoping of the current design. Given the modular nature of the instrument, it is therefore natural to consider a reduced number of modules with a correspondingly smaller wavelength range. However it is important that the simultaneous wavelength coverage is as large as possible, encompassing both optical and near infrared, for the following reasons.
\begin{itemize}
\item The scientific importance of simultaneous optical and near infrared observations has been clearly demonstrated by X-Shooter at the VLT and has been emphasized for HIRES at the E-ELT all along. However, the real strength of HIRES is the combination of the large wavelength coverage and high spectral resolution with the large photon-collecting capacity. HIRES would also be the only instrument at any ELT that provides polarimetric information. 
\item HIRES is intended to be a general purpose high resolution spectrograph for the community of ESO member states and must be ready to meet the scientific challenges of the next decades. This requires a large as possible wavelength range, encompassing at least both the optical and the near-IR.
\item The E-ELT telescope time will be too valuable not to exploit it to the largest possible extent. High-resolution spectra with an X-Shooter like wavelength range would have an enormous legacy value for the future community.
\end{itemize}

\acknowledgments % equivalent to \section*{ACKNOWLEDGMENTS}       

This work was supported from the Italian National Institute for Astrophysics (Istituto Nazionale Italiano di Astrofisica, INAF).
RM , DB, CH, MF, XS, DQ and MGH acknowledge support from the UK Science and Technology Facilities Council (STFC). MGH is supported by the ERC Advanced grant Emergence-32056.
This work was supported by Funda\c{c}\~ao para a Ci\^encia e a Tecnologia (FCT, Portugal), project ref. PTDC/FIS-AST/1526/2014, through national funds and by FEDER through COMPETE2020 (ref. POCI-01-0145-FEDER-016886), as well as through grant UID/FIS/04434/2013 (POCI-01-0145-FEDER-007672).
P.F. and N.C.S. also acknowledge the support from FCT through Investigador FCT contracts of reference IF/01037/2013, IF/00169/2012, and IF/00028/2014, respectively, and POPH/FSE (EC) by FEDER funding through the program ``Programa Operacional de Factores de Competitividade - COMPETE''. P.F. further acknowledge support from FCT in the form of exploratory projects of reference IF/01037/2013CP1191/CT0001 and IF/00028/2014/CP1215/CT0002.
PJA acknowledges financial support from AYA2011-30147-C03-01 and AYA2014-54348-C3-1-R by MINECO/Spain, partially supported by FEDER funds/EU.
Research activities of the Board of Stellar Astronomy, at the Federal  University of Rio Grande do Norte are supported by continuous grant of CNPq, FAPERN and CAPES brazilian agencies and  by the INCT INEspa\c{c}o.
E.D.M and V.Zh.A. also acknowledge the support from the FCT (Portugal) in the form of the grants SFRH/BPD/76606/2011 and SFRH/BPD/70574/2010, respectively.
% References
\bibliography{biblio} % bibliography data in report.bib
\bibliographystyle{spiebib} % makes bibtex use spiebib.bst

\end{document}